\documentclass[journal]{IEEEtranTIE}

\usepackage{graphicx}
\usepackage{cite}
\usepackage{picinpar}
\usepackage{amsmath}
\usepackage{url}
\usepackage{flushend}
\usepackage[latin1]{inputenc}
\usepackage{colortbl}
\usepackage{soul}
\usepackage{multirow}
\usepackage{pifont}
\usepackage{color}
\usepackage{alltt}
\usepackage[hidelinks]{hyperref}
\usepackage{enumerate}
\usepackage{siunitx}
\usepackage{pbox}
\usepackage{tikz}
\usepackage[font={footnotesize,sf}]{caption}
\usepackage[font={footnotesize,sf},skip=3pt]{subcaption} 
\usetikzlibrary{arrows.meta,positioning}
\usepackage{comment}

\definecolor{bostonuniversityred}{rgb}{0.8, 0.0, 0.0}
\definecolor{limegreen}{rgb}{0.2, 0.8, 0.2}
\definecolor{forestgreen}{rgb}{0.13, 0.55, 0.13}
\definecolor{greenhtml}{rgb}{0.0, 0.5, 0.0}
\definecolor{filln}{rgb}{0.8, 0.9, 1.0}
\definecolor{linen}{rgb}{0.0, 0.1, 0.2}
\definecolor{fillt}{rgb}{1.0, 0.85, 0.7}
\definecolor{linet}{rgb}{0.3, 0.15, 0.0}
\definecolor{filld}{rgb}{1.0, 0.7, 1.0}
\definecolor{lined}{rgb}{0.3, 0.0, 0.15}
\definecolor{fillg}{rgb}{0.77, 0.9, 0.77}
\definecolor{lineg}{rgb}{0.15, 0.2, 0.15}

\newcommand \auxdRelg	{ \mathcal R^{\prime \, *}_\mathrm{g} }
\newcommand \auxdRelgz	{ \mathcal R^{\prime \, *}_\mathrm{g0} }
\newcommand \auxRelc	{ \mathcal R^{ \, *}_\mathrm{c} }
\newcommand \auxRelcz	{ \mathcal R^{ \, *}_\mathrm{c0} }
\newcommand \auxRelg	{ \mathcal R^{ \, *}_\mathrm{g} }
\newcommand \auxRelgz	{ \mathcal R^{ \, *}_\mathrm{g0} }
\newcommand \ddzref	    { \ddot{\pos}_\mathrm{d} }
\newcommand \dlambdaref { \dot \lambda_\mathrm{d}}
\newcommand \dzref		{ \dot{\pos}_\mathrm{d} }
\newcommand \I		{ I }
\newcommand \icoil		{ i }
\newcommand \iref		{ i_\mathrm{d} }
\newcommand \lambdaref	{ \lambda_\mathrm{d} }
\newcommand \lambdasat	{ \lambda_\mathrm{sat} }
\newcommand \lfrac[2]	{ #1 / #2 }
\newcommand \lsat		{ \lambda^*_\mathrm{sat} }
\newcommand \param		{ p }
\newcommand \Parmag		{ \Gamma_\mathrm{mag} }
\newcommand \Parop		{ \Gamma_\mathrm{e} }
\newcommand \Paropn[1]	{ \Gamma_{#1} }
\newcommand \pos        { \theta }
\newcommand \Relc		{ \mathcal R_\mathrm{c} }
\newcommand \Relg		{ \mathcal R_\mathrm{g} }
\newcommand \tc			{ t_\mathrm c }
\newcommand \tf			{ t_\mathrm{f} }
\newcommand \uaudio		{ \upsilon_\mathrm{audio} }
\newcommand \ucoil		{ u }
\newcommand \vcoil		{ u }
\newcommand \vcoilref	{ u_\mathrm{d} }
\newcommand \vel        { \dot \theta }
\newcommand \zc		    { \pos_\mathrm{c} }
\newcommand \zf			{ \pos_\mathrm{f} }
\newcommand \zmax		{ \pos_\mathrm{max} }

\newcommand \zNC        { \pos_\mathrm{NC} }
\newcommand \zNO        { \pos_\mathrm{NO} }
\newcommand \zref		{ \pos_\mathrm{d} }

\begin{document}

\title{\huge{An Audio-Based Iterative Controller for\\Soft Landing of Electromechanical Relays}}

\author{
	
	Eloy Serrano-Seco,
	Edgar Ramirez-Laboreo,
	Eduardo Moya-Lasheras,
	and Carlos Sagues

	\thanks{
		This work was supported in part by the Spanish Government/EU, under projects RTC-2017-5965-6, PGC2018-098719-B-I00, PID2021-124137OB-I00, TED2021-130224B-I00, and CPP2021-008938, in part by the Arag\'on Government/EU, under project DGA{\_}FSE T45{\_}20R, and in part by the European Union - Next Generation EU. (Corresponding author: \textit{Eloy Serrano-Seco}.)
		
		The authors are with the Departamento de Informatica e Ingenieria de Sistemas (DIIS) and the Instituto de Investigacion en Ingenieria de Aragon (I3A), Universidad de Zaragoza, 50018 Zaragoza, Spain (e-mail: eserranoseco@unizar.es; ramirlab@unizar.es; emoya@unizar.es, csagues@unizar.es). 
	}
     \thanks{\textcolor{red}{This is the accepted version of the manuscript: E. Serrano-Seco, E.~Ramirez-Laboreo, E.~Moya-Lasheras and C. Sagues, ``An Audio-Based Iterative Controller for Soft Landing of Electromechanical Relays,'' in \textit{IEEE Transactions on Industrial Electronics}, vol. 70, no. 12, pp. 12730-12738, Dec. 2023, doi: 10.1109/TIE.2022.3231254. 
    \textbf{Please cite the publisher's version}. For the publisher's version and full citation details see: \protect\url{https://doi.org/10.1109/TIE.2022.3231254}. 
    }}
}

\maketitle
	
\begin{abstract}
    Electromechanical relays and contactors suffer from strong collisions at the end of the switching operations. This causes several undesirable phenomena, such as clicking, mechanical wear and contact bounce. Thus, there is great interest in mitigating these switching impacts while keeping the advantageous features of these devices. This paper proposes a complete control strategy for soft landing. The control structure includes three main components. The first one is a real-time flux-tracking feedback controller, which presents several advantages over voltage or current control. The second one is a feedforward controller, which computes the flux reference signal based on a proposed dynamical model and the desired position trajectory for the switching operations. Lastly, the third control component is a learning-type run-to-run adaptation law that iteratively adapts the model parameters based on an audio signal. It exploits the repetitive nature of these devices in order to circumvent modeling discrepancies due to unit-to-unit variability or small changes between operations. The effectiveness of the proposed control is demonstrated through various experiments.
\end{abstract}


\section{Introduction}\label{sec:intro}

\IEEEPARstart{E}{lectromechanical} relays are used in many applications, e.g. drive by wire~\cite{Naidu2010}, induction heating~\cite{Acero2010}, battery charging~\cite{haghbin2013} or wireless power transfer~\cite{beh2013}. They offer many advantages compared to solid state switches: they are generally cheaper and more efficient, are able to conduct and block current in both directions, provide electrical isolation between the activation circuit and the power terminals, and have a simple activation mode. Additionally, whereas alternative semiconductor devices can only provide single-pole single-throw arrangements, relays can be designed as multiple-pole and multiple-throw, which is really useful in various applications. 

Electromechanical relays, however, also present some drawbacks. One of their most known and undesirable problems is the strong switching impacts between the movable and fixed components, which cause wear, contact bouncing and an acoustic noise that is undesirable in certain domestic applications. Additionally, bounces may provoke contact welding~\cite{Barkan1967} or arcing that exacerbates the erosion~\cite{mcbride1991}. These problems lead to a reduction in the service life of these devices and the equipment in which they are embedded, and limit the range of applications in which electromechanical relays are the best switching solution. Therefore, there is a great interest in mitigating the switching impacts of electromechanical relays while keeping their advantages over solid-state alternatives.

The state-of-the-art soft-landing approaches found in the literature focus almost solely on reducing the contact bounces and their duration. In addition, the vast majority of works focus on contactors, which are electromechanical relays for high power switching. 
In 1996, Davies et al.~\cite{davies1996} presented one of the first control-oriented approaches for contact bounce reduction. The proposed open-loop controller was however not robust because it did not have the ability to self-adapt to changing conditions or perturbations. Then, based on this work, improvements were made to increase robustness. For instance, a current controller was proposed in~\cite{demoraes2008}, which detects the start of the closing process and accordingly modifies the coil energization. As a different approach to increase robustness, a run-to-run control was proposed in~\cite{ramirez-laboreo2017}, which iteratively adapts voltage pulse durations. The main limitation of this approach is its sensitivity to resistance variations due to temperature changes. Another recent work~\cite{tang2021} proposed a current tracking solution to reduce contact bounce, whose excitation time is iteratively adapted with a fuzzy controller. The current tracking controller circumvents the variability due to temperature changes, but the current reference is constant, which theoretically cannot guarantee perfect soft landing~\cite{Moya-Lasheras2020tcst}.

In parallel, displacement or position feedback control has also been investigated. In 1999, Carse et al.~\cite{carse1999} showed via simulation that a simple fuzzy controller can reduce contact bouncing. The main obstacle of this approach is the need of accurate measurements of the mechanism position. In most cases, position sensors are too expensive, or the movable parts are not accessible. As workarounds, some works present position estimators, e.g. model-based~\cite{wang2010} or neural networks~\cite{tang2021}. However, these estimators are very sensitive to discrepancies between the dynamics of different relays.

In short, the soft-landing control problem is still not satisfactorily solved for electromechanical relays. Some works propose position feedback controllers~{\cite{carse1999,wang2010,Espinosa2007}} that may be used for position tracking and ultimately soft landing. However, they rely on position sensors---which cannot be implemented in most applications---or estimators---which are too computationally expensive, inaccurate or sensitive to modeling and measurement errors. As a possible solution, other works present open-loop or iterative controllers that do not rely on position measurements or estimates~{\cite{davies1996,demoraes2008,ramirez-laboreo2017,tang2021,Lin2013,tang2022}}. However, they focus on reducing impact bounces, which can be counterproductive for the soft landing objective (e.g., strong switching forces may reduce the number of bounces but increase the impacts). Furthermore, most solutions are sensitive to temperature variations~{\cite{davies1996,ramirez-laboreo2017}} or modeling discrepancies~{\cite{davies1996,wang2010,tang2021,tang2022}}.

This paper presents a new control strategy for impact reduction in electromechanical relays. The main contribution is the combination of three distinct components: a real-time flux-tracking controller, a flatness-based feedforward controller, and a learning-type run-to-run adaptation law. The flux-tracking controller relies on a real-time flux estimator which is fed with easily measurable electrical signals, i.e., voltage and current. The main advantage of a flux controller over a voltage or current controller is that this approach depends on fewer parameters, which are a main source of uncertainty and control variability.

The feedforward controller, which generates the desired flux trajectory, is based on a dynamical model also presented in the paper. The main contribution of this model is that it separates the motion of the mechanism into different stages, taking into account the deformation of several components of the device. The flatness property of the model provides a method to obtain the flux trajectory used by the flux-tracking controller from a desired position trajectory for the armature, which is in turn designed considering the different phases of the motion.
Lastly, the run-to-run adaptation law iteratively adapts the model parameters---required by the feedforward controller---making the overall control more robust to any type of parameter variability. The adaptation law is formulated as an optimization problem that iteratively modifies the parameters of the model in order to maximize the system performance. This performance is computed based on a microphone signal of the noise generated during the switchings. Thus, instead of focusing on reducing contact bounce---as is most common for relays---our soft-landing approach consists in reducing impact sounds and, indirectly, impact velocities.\looseness-1

\section{Model}\label{sec:model}

\begin{figure}[t]
	\centering
	\begin{subfigure}[b]{0.5\linewidth}
		\centering
		\includegraphics[width=0.95\textwidth]{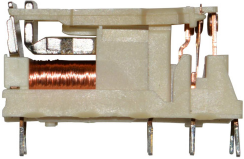}
	\hspace*{\stretch{1}}%
    \caption{\label{fig:relay_photo}}
    \end{subfigure}%
	\begin{subfigure}[b]{0.5\linewidth}
		\centering
		\includegraphics[width=0.95\textwidth]{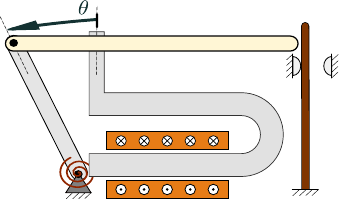}
	\hspace*{\stretch{2}}%
	\vspace{0.5cm}
    \caption{\label{fig:relay_geom}}
    \end{subfigure}%
	\caption{Electromechanical relay. (a) Photo. (b) Schematic diagram.}
	\label{fig:relay}
\end{figure}

The electromechanical relay used in this work is shown in Fig.~\ref{fig:relay}. It is a single-pole double-throw relay with a U-shaped magnetic core and a moving core, or armature. There are two distinct operations depending on the direction of the armature motion: making and breaking, when the coil energizes or de-energizes, respectively. The armature is connected with the movable electrical contact via a plastic component. In the making operations, the contact is pushed toward the normally open position, whereas in the breaking operations, the contact is pulled toward the normally closed position.

In order to design the controller, it is necessary to characterize the dynamics of the system. For clarity, the model is divided into two interconnected systems. One of them is the electromagnetic system, which is governed by two main equations. The first one is the electrical circuit equation,
\begin{align}\label{eq:el_circuit}
    \vcoil = R\, \icoil + \dot \lambda, 
\end{align}
where $\vcoil$, $R$, $\icoil$ and $\lambda$ are the voltage between the coil terminals, the coil internal resistance, the coil current, and the magnetic flux linkage, respectively. The second equation refers to the magnetic equivalent circuit,
\begin{equation}\label{eq:el_mec}
    N^2\,\icoil = \big(\Relc(\lambda) + \Relg(\pos)\big) \, \lambda,
\end{equation}
where $N$ is the number of coil turns, and $\Relc$ and $\Relg$ are respectively the magnetic reluctances of the iron core and the air gap. These reluctances can be modeled as functions of the flux, $\lambda$, and the angular position of the armature, $\pos$, taking into account the magnetic saturation and flux fringing phenomena, respectively~\cite{moya-lasheras2021mech}. In order to simplify these expressions and the consequent model-based controller, the following auxiliary functions are defined as scaled versions of the reluctances,
\begin{align}
	\auxRelc(\lambda) &= \frac{\Relc(\lambda)}{N^2} = \frac{\auxRelcz}{1-|\lambda|/\lambdasat}, \label{eq:relc}\\
	\auxRelg(\pos) &= \frac{\Relg(\pos)}{N^2} = \auxRelgz + \frac{\auxdRelgz \, \pos}{1 + \kappa_1\, \pos \, \ln(\kappa_2/\pos)}, \label{eq:relg}
\end{align}
being $\auxRelcz$, $\lambdasat$, $\auxRelgz$, $\auxdRelgz$, $\kappa_1$ and $\kappa_2$ positive constants. Using \eqref{eq:el_circuit}--\eqref{eq:relg}, the flux linkage dynamics is derived as
\begin{equation}\label{eq:dflux}
    \dot \lambda = - R \, \big( \auxRelc(\lambda) + \auxRelg(\pos) \big) \,\lambda + \ucoil.
\end{equation}

\begin{figure}
     \subcaptionbox{\label{fig:mech_1}}
     [0.33\linewidth]{\includegraphics[scale=1.25]{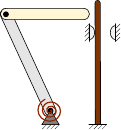}}%
     \subcaptionbox{\label{fig:mech_2}}
     [0.33\linewidth]{\includegraphics[scale=1.25]{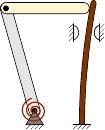}}%
     \subcaptionbox{\label{fig:mech_3}}
     [0.33\linewidth]{\includegraphics[scale=1.25]{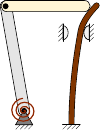}}
 	 \caption{Schematic diagram of the different stages during the armature movement. (a) Only the armature and the plastic component are in motion. (b) The plastic component pushes and deforms the moving contact. (c) The moving contact is already touching the normally open contact, but can still be deformed.}
 	\label{fig:mech}
 \end{figure}

Regarding the mechanical system, it must be noted that, despite its small size and apparent simplicity, the relay mechanism shown in Fig.~\ref{fig:relay} is in fact rather complex. There are actually three moving components: the armature, the plastic part at the top and the moving contact. Due to mechanical play, however, these three components do not always move together.
When at rest, the armature remains at a position defined by $\pos=\zmax$. In the first phase of the motion, the armature and the plastic component move solidly without making contact with the moving contact, which is touching the normally closed (NC) contact (see Fig.~\ref{fig:mech_1}). Then, when the armature reaches the position $\pos = \zNC$, the plastic part touches the moving contact and the three components start to move solidly (Fig.~\ref{fig:mech_2}). There is an armature position, $\pos=\zNO$, at which the moving contact reaches the normally open (NO) contact.
However, due to the geometry and elasticity of this latter component, the armature and the plastic component can still continue to move forward by deforming the moving contact (Fig.~\ref{fig:mech_3}).

Some assumptions are taken in order to characterize the motion of this mechanism in a simple but detailed way. Firstly, it is assumed that all the mass is concentrated in the moving armature, which is by far the heaviest component of the three previously mentioned. Furthermore, it is also assumed that the three phases of the motion described above are the same regardless of whether the armature is moving in one direction or the other. That is, it is assumed that the position of the three components can be unambiguously defined based solely on the position of the armature $\theta$. Under these simplifications, the motion of the mechanism can be described by Newton's second law,
 \begin{equation}\label{eq:newton2}
 	\I \, \ddot \pos = \Parmag(\pos,\lambda) + \Parop(\pos) + c \, \vel,
 \end{equation}
 where $\I$ is the armature moment of inertia, $\Parmag$ and $\Parop$ are respectively the magnetic and elastic torques, and $c$ is a damping coefficient. The magnetic torque depends on the derivative of the gap reluctance and the flux~\cite{ramirez-laboreo2016} or, equivalently, on the derivative of the auxiliary function $\auxRelg$ and the flux linkage,
 \begin{align}\label{eq:Fmag}
 	\Parmag(\pos,\lambda) = - \frac{\auxdRelg(\pos) \, \lambda^2}{2} , &  & \auxdRelg(\pos) = \frac{\partial \auxRelg(\pos)}{\partial \pos}.
 \end{align}
 The elastic torque, on the other hand, is modeled as the following piecewise function,
 \begin{equation}\label{eq:opposite_force}
 \Parop(\pos)= 
   \begin{cases}
       \Paropn{e,1}, & \mbox{if } \pos > \zNC \\
       \Paropn{e,2}, & \mbox{if } \zNO \leq \pos  \leq \zNC \\
       \Paropn{e,3}, & \mbox{if } \pos < \zNO
       \end{cases},
 \end{equation}
 which defines a specific expression for each stage of the motion. In the first stage (Fig.~\ref{fig:mech_1}), there is a single elastic torque that opposes the magnetic force and tends to keep the armature position at $\theta = \zmax$. It is modeled as an ideal torsion spring,
\begin{equation}
    \Paropn{e,1}  = k_1 \, (\pos_\mathrm{max}  -  \pos),
\end{equation}
where $k_1$ is the stiffness constant. In the second stage (Fig.~\ref{fig:mech_2}), the elastic torque is increased due to the deformation of the contact. According to the Euler--Bernoulli beam theory, and assuming small deformation and angles, this can be approximated as an additional proportional elastic torque,
\begin{equation}\label{eq:torque_2}
    \Paropn{e,2} = k_1 \, (\zmax  -  \pos) + k_2\,(\zNC-\pos),
\end{equation}
where $k_2$ is the stiffness constant due to the moving contact deformation. Lastly, in the third stage (Fig.~\ref{fig:mech_3}), the moving contact has reached its limit, but part of it still deforms. Following the same reasoning as in~\eqref{eq:torque_2}, the total torque is
\begin{equation}\label{eq:torque_3}
     \Paropn{e,3} = k_1 \, (\pos_\mathrm{max}  -  \pos)  + k_2 \,  (\zNC  -  \zNO) + k_3 \, (\zNO  -  \pos),
\end{equation}
where $k_3$ is the stiffness constant due to the part of the moving contact that is still warping.

\section{Control}\label{sec:control}

\begin{figure*}[t]
	\centering
	\hspace{3mm}
	\def\sumoffset{1mm}
	\def\sumoffsetaux{1.5mm}
	\def\nodex{10mm}
	\def\nodey{12mm}
	\def\arrowsep{3mm}
	\def\lwt{0.3mm}
	\def\lwn{0.4mm}
	\def\lwd{0.5mm}
	\hspace{-8mm}
	\vspace{1mm}
	\begin{tikzpicture}[
		node distance = \nodey and \nodex,
		box/.style = {draw, minimum height=6mm, minimum width=12mm, align=center},
		bigbox/.style = {draw, dashed,draw=linen, minimum height=13mm, minimum width=24mm, align=center},
		sum/.style = {circle, draw, node contents={}},
		>={Stealth[width=2mm,length=3mm]}
		]
		\node (ref) [] {$\zref(t)$};
		\node (ff) [box, fill=fillg, draw=lineg, right=of ref] {Feedforward \\ controller};
		\node (traj) [box, fill=filld, draw=lined, left=of ref] {Position\\trajectory\\design};
		\node (s1) [sum, draw=linet, right=of ff,xshift=8mm, anchor=center];
		\node (ctrl) [box, fill=fillt, draw=linet, right=of s1,xshift=3mm] {Flux-tracking controller};
		\coordinate[right=of ctrl] (c2);
		\node (plant) [box, right=of c2] {Plant};
		\coordinate[below=of plant.center, yshift=-0.5*\arrowsep] (c5);
		\node (est) [box, fill=fillt, draw=linet, below=of ctrl.center, anchor=center] {Flux estimator};
		\coordinate[right=of est,xshift=-\nodex,yshift=0.5*\arrowsep] (c1_est);
        \coordinate[right=of est,xshift=-\nodex,yshift=-0.5*\arrowsep] (c2_est);
		\node (c6) [below=of s1.center, anchor=center] {};
		\node (estaux) [left=of est, xshift=6mm, yshift=1.5mm] {};
		\coordinate[above=of plant.center] (c7);
		\coordinate[above=of c7] (c8);
		\coordinate[above=of c8] (c9);
		\coordinate[above=of ff.center] (c10);
		\node (opt) [box, fill=filln, draw=linen, above=of ctrl.center, anchor=center] {Run-to-run\\adaptation law};
		\node (hold) [box, fill=filln, draw=linen, above=of s1.center, anchor=center] {Hold};
		\draw[->,draw=linet,line width=\lwt] (ff) -- node[above, xshift=-1mm] {$\lambdaref^{n}(t)$} (s1);
		\draw[->,draw=linet,line width=\lwt] (s1) -- (ctrl);
		\draw[->,draw=linet,line width=\lwt] (ctrl) -- node[above] {$\ucoil^{n}(t)$} (plant);
		\draw[->,draw=linet,line width=\lwt] (est) -| (s1);
		\draw[->,draw=linet,line width=\lwt] (c2) |- (c1_est);
		\draw[-,draw=linet,line width=\lwt] (plant) -- node[right] {$ \icoil^{n}(t) $} (c5);
		\draw[->,draw=linet,line width=\lwt] (c5) -- node[above] {$ $} (c2_est);
		\draw[-,draw=linet,line width=\lwt] (est) -- node[above,xshift=-3mm] {$\hat \lambda^{n}(t)$} (c6);
		\draw[dashed,draw=linen,line width=\lwn] (plant) -- node[right] {$\uaudio^{n}(t)$} (c7) -- (opt);
		\draw[dashed,draw=linen,line width=\lwn] (opt) -- node[above] {$\param^{n+1}$} (hold);
		\draw[dashed,draw=linen,line width=\lwn] (hold) -- node[above] {$\param^{n}$} (c10);
		\draw[dashed,draw=linen,->,line width=\lwn] (c10) -| (ff);
		\draw[draw=linet,->,line width=\lwt] (ref) -- (ff);
		\draw[dotted,draw=lined,->,line width=\lwd] (traj) -- (ref);
		\footnotesize
		\node (s1+) [above left=of s1.center, xshift=\nodex-\sumoffsetaux, yshift=-\nodey+\sumoffset] {\color{linet}\textbf{$\mathbf +$}};
		\node (s1-) [below left=of s1.center, xshift=\nodex-\sumoffset, yshift=\nodey-\sumoffsetaux] {\color{linet}$\mathbf -$};
	\end{tikzpicture}
	\caption{Control diagram. The superscript $n$ is used to denote the variables of the $n$th operation. The inner loop (blocks in orange) is the real-time flux-tracking controller. The flux linkage reference $\lambdaref$ is provided in real time by the feedforward controller (in green) which is fed with the desired position trajectory $\zref$. The run-to-run adaptation law (in blue) uses the audio signal $v_\mathrm{audio}$ to update the parameter set $p$ of the feedforward controller only once per operation.}
	\label{fig:ctrl_diag}
\end{figure*}
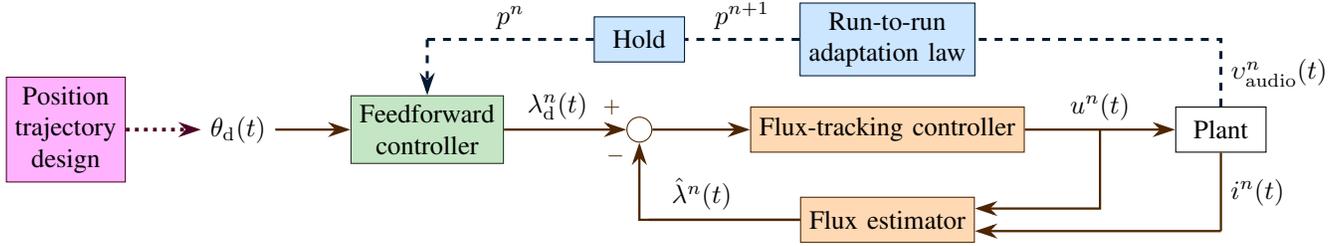

The proposed controller is schematized in Fig.~\ref{fig:ctrl_diag}. It can be interpreted as a cascade controller with two feedback loops. The inner loop is an online flux-tracking controller which is fed with the estimated flux linkage $\hat \lambda$. On the other hand, the outer loop is a learning-type controller that iteratively adapts the reference flux linkage $\lambdaref$, based on the model parameters $\param$. This reference flux is obtained through a feedforward controller based on the desired position trajectory $\zref$.

\subsection{Position trajectory design}\label{sec:pos_ref}

\begin{figure}
	\includegraphics{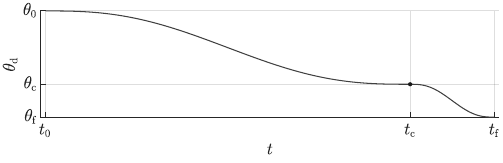}
	\caption{Desired position trajectory based on two concatenated $5$th-degree polynomial trajectories.}
	\label{fig:z_ref}
\end{figure}

Firstly, the desired position trajectory is defined, considering that the objective is to reduce the impact noise. Most commonly, soft-landing trajectories are defined such that the final velocity and acceleration are minimized~\cite{Moya-Lasheras2020tcst}. However, for the studied relay, there are two main impact events that cause noise and wear: the stroke ends of the moving contact and of the armature. Taking this into account, this paper proposes a concatenation of two soft-landing trajectories. Overall, the combined position trajectory has the following boundary conditions:
\begin{align}
    \zref(t_0) = \pos_0, && \dzref(t_0) = 0, && \ddzref(t_0) = 0, \label{eq:bound1} \\
    \zref(\tc) = \zc, && \dzref(\tc) = 0, && \ddzref(\tc) = 0, \label{eq:bound2} \\
    \zref(\tf) = \zf, && \dzref(\tf) = 0, && \ddzref(\tf) = 0, \label{eq:bound3}
\end{align}
where $t_0$, $\tc$ and $\tf$ are the initial, contact and final user-defined instants and $\pos_0$, $\zc$ and $\zf$ are the initial, contact and final positions, which respectively correspond to $\zmax$, $\zNO$ and $0$ for the making (i.e. closing) operations, or to $0$, $\zNC$ and $\zmax$ for the breaking (i.e. opening) operations. The two concatenating trajectories are designed as $5$th-degree polynomials. The first one defines the trajectory for $t \in \left[ t_0, \; \tc \right]$, while the second one applies to the interval $t \in \left[ \tc, \; \tf \right]$. Thus, the boundary conditions~{\eqref{eq:bound1}} and~{\eqref{eq:bound2}} are used to solve the six polynomial coefficients corresponding to the trajectory of the first interval, whereas~{\eqref{eq:bound2}} and~{\eqref{eq:bound3}} are used to solve the six polynomial coefficients for the second interval. The general form of the designed position trajectory is represented in Fig.~\ref{fig:z_ref}.

\subsection{Feedforward controller}\label{sec:flatness}

The feedforward controller of this paper exploits the differential flatness property of the model. In short, an $n$th-order system is differentially flat if the $n$th derivative of the output is the first one where the input appears explicitly~\cite{levine2011}.
In this case, it can be shown that the angular position is a flat output of the dynamical model presented in the previous section. 
As a consequence, the desired flux linkage $\lambdaref$ can be calculated from the desired position $\zref$ and its derivatives as
\begin{equation}\label{eq:lambda_flatness}
	\lambdaref = \sqrt{\lfrac{2\,(\Parop(\zref) - c\,\dzref - \I \, \ddzref)}{\auxdRelg(\zref)}}.
\end{equation}
This expression, which results from {\eqref{eq:newton2}}--{\eqref{eq:torque_3}}, depends on the parameter vector $p$, given by
\begin{equation}\label{eq:param0}
    \param = \left[\begin{array}{cccccccc}
        \I & k_1 & k_2 & k_3 & c & \auxdRelgz & \kappa_1 & \kappa_2 
    \end{array}\right].
\end{equation}
The desired current $\iref$ may also be obtained from {\eqref{eq:el_mec}}--{\eqref{eq:relg}} as
\begin{equation}\label{eq:i_flatness}
	\iref = \big(\auxRelc(\lambdaref) + \auxRelg(\zref)\big) \, \lambdaref,
\end{equation}
where $\lambdaref$ is given by~{\eqref{eq:lambda_flatness}}. This expression, which is slightly more complex than the previous one, depends on three additional parameters: $\auxRelcz$, $\lsat$ and $\auxRelgz$.
Finally, the desired voltage $\vcoilref$ might also be computed using {\eqref{eq:el_circuit}} as
\begin{equation}\label{eq:u_flatness}
	\vcoilref = \dlambdaref + R \, \iref,
\end{equation}
where $\dlambdaref$ is the time derivative of {\eqref{eq:lambda_flatness}} and $\iref$ is given by {\eqref{eq:i_flatness}}. Again, this expression depends on an additional parameter: the resistance $R$.

At this point, it is obvious that if all the parameters were known, it would be practically the same to actuate using voltage, current or flux linkage. However, if there is any uncertainty with the parameter values, in terms of robustness it is clearly advantageous to control the flux linkage, whose desired value depends on fewer parameters than the other variables.
For this reason, the feedforward controller presented in this paper defines the desired trajectory of the flux linkage. The following sections explain how this desired trajectory is tracked and how the uncertainty of the parameter vector $p$ is managed.

\subsection{Flux-tracking controller}\label{sec:flux_control}

The flux linkage $\lambda$ is a variable that cannot be easily measured. However, it is possible to obtain an estimate based only on measurements of the electrical variables, i.e., voltage and current. In this work, we estimate $\lambda$ by means of the reset estimator presented and discussed in~{\cite{Electromagnetic-Estimation}}.
An important issue with this estimator is the resistance value, which is required by the algorithm but varies with temperature. Our proposal is to estimate this parameter just before each operation. This can be achieved through Ohm's law, simply by applying a small constant voltage and measuring the steady-state electrical current. Note that the thermal dynamics of the device is much slower than the electromechanical one, so it is reasonable to assume that $R$ remains constant during each operation.

Once we have an estimate of the flux linkage, we can proceed to the design of the tracking controller. Our proposal is based only on the dynamics described by {\eqref{eq:dflux}}. In that equation, the key is to note that $R \, \big( \auxRelc(\lambda) + \auxRelg(\pos) \big)>0$ for all $\lambda$ and $\theta$ [see {\eqref{eq:relc}} and {\eqref{eq:relg}}]. Thus, the flux linkage dynamics can be regarded as a first-order linear time-variant system,
\begin{equation}\label{eq:mag_circuit}
	\dot \lambda(t) = - a(t) \,\lambda(t) + \ucoil(t),
\end{equation}
where $a(t) = R \left( \auxRelc\left(\lambda\left(t\right)\right) + \auxRelg\left(\pos\left(t\right)\right) \right)$ is strictly positive, which guarantees the system stability. Then, to control $\lambda$ we use a proportional-integral (PI) controller in parallel form,
\begin{equation}\label{eq:PI}
    u(t) = K_\mathrm{p}\,e(t) + K_\mathrm{i} \int_0^t e(\tau)\, \mathrm{d}\tau, \quad e(t) = \lambda_\mathrm{d}(t)-\hat\lambda(t),
\end{equation}
where $\hat\lambda$ is the estimate of the flux linkage. Given that {\eqref{eq:mag_circuit}} is strictly stable, it can be easily shown that the closed-loop system is also stable for any positive $K_\mathrm{p}$ and $K_\mathrm{i}$ (see proof in Appendix~{\ref{appendix:stability}}).

\subsection{Run-to-run adaptation law}\label{sec:r2r}
The performance of the flux-tracking controller depends on the accuracy of the model, which is directly used by the feedforward term. There are various sources of modeling discrepancies, such as variability between units due to large manufacturing tolerances, or variability between operations of each unit due to mechanical wear. Thus, to increase the control robustness, a learning-type run-to-run adaptation law is incorporated to iteratively adapt the model parameters.

Ideally, the reference to this control loop would be the angular position, but position sensing is usually unfeasible---the moving components may not be physically reachable---or unaffordable---position sensors are much more expensive than the relays themselves. Nevertheless, it is possible to obtain auxiliary measurements for evaluating the performance of the inner-loop flux-tracking controller, such as the impact sound or the bouncing duration. As stated in the introduction, many control proposals for relays are aimed at reducing contact bounce, but another undesirable problem, the acoustic noise, is often neglected. Both undesirable problems are not entirely related, i.e. minimizing contact bouncing does not necessarily imply that the acoustic noise is also reduced. Furthermore, reducing the noise involves further difficulties, e.g. the control should be applied to the entire armature stroke, even after the electrical contact has reached the final position. For these reasons, this paper proposes to measure an audio signal, $\uaudio$, captured with a microphone during each switching operation. Then, for each iteration, a cost $J$ is computed from this signal in order to evaluate the operation. This paper proposes
\begin{equation}\label{eq:cost}
    J = \int_{t_\mathrm{0}}^{t_0+\Delta t} {\uaudio}^2(t) \, \mathrm d t,
\end{equation}
where $\Delta t$ should be large enough to capture all the acoustic noise generated during the switching.

Then, the control task is formulated as an optimization problem that searches the space of the model parameters that are involved in the feedforward term.
The parameter vector $p$ [see {\eqref{eq:param0}}] is modified in each switching operation in order to minimize the cost $J$ calculated from the audio signal. Note that this is a black-box optimization problem, because the function that relates the model parameters to the cost is unknown. As such, the convergence to an optimal performance point depends on the shape of the cost function, the dimensionality of the search space and the convergence properties of the optimization algorithm selected for the task. Note that, in this aspect, the flux-based feedforward controller is also advantageous over the voltage- or current-based ones because it depends on fewer parameters. The optimization method used in this work is based on the Nelder--Mead method~\cite{nelder1965}, which is a widely used and studied direct-search optimization method~\cite{Kolda2003a}. Although it was designed for offline optimization of deterministic functions, it has been adapted for online optimization of stochastic functions in several works. In these cases, however, there is no theoretical proof of convergence to an optimal point, so it requires simulated or experimental validation. Specifically, this paper uses a modified version that has been previously proposed and validated for the control of a different class of switch-type actuators~\cite{Moya-Lasheras2020tmech}.

As a final remark, it should be noted that the goal of the adaptation law is not to estimate the model parameters, but to reduce the acoustic noise due to switching impacts. Thus, regardless of the optimization method, there is no guarantee that the adapted values converge to the true ones, but simply to a set of values that minimize the noise.

\section{Experimental validation}\label{sec:exp}

\subsection{Prototype and experimental setup}\label{sec:setup}

\begin{figure}
\centering
	\includegraphics[width=0.9\linewidth]{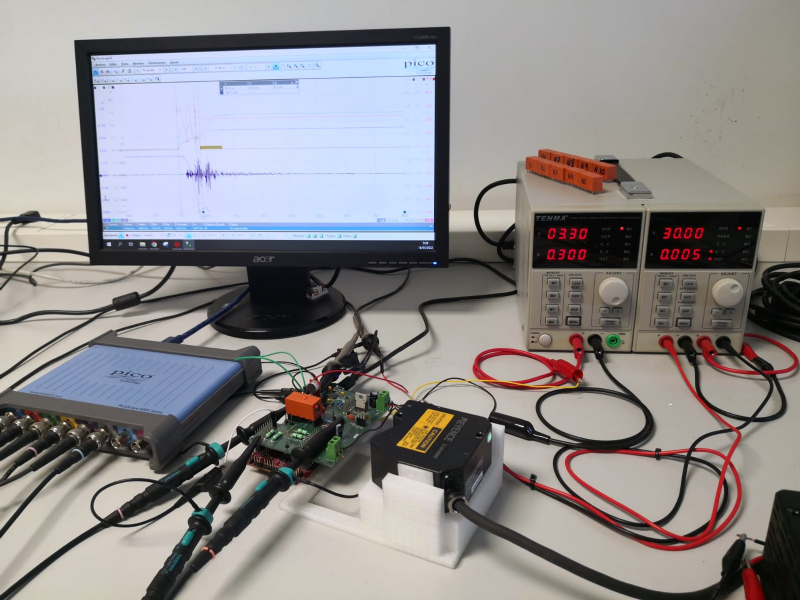}
	\caption{Experimental setup.}
	\label{fig:environment}
\end{figure}

\begin{figure}
\centering
	\includegraphics[width=0.9\linewidth]{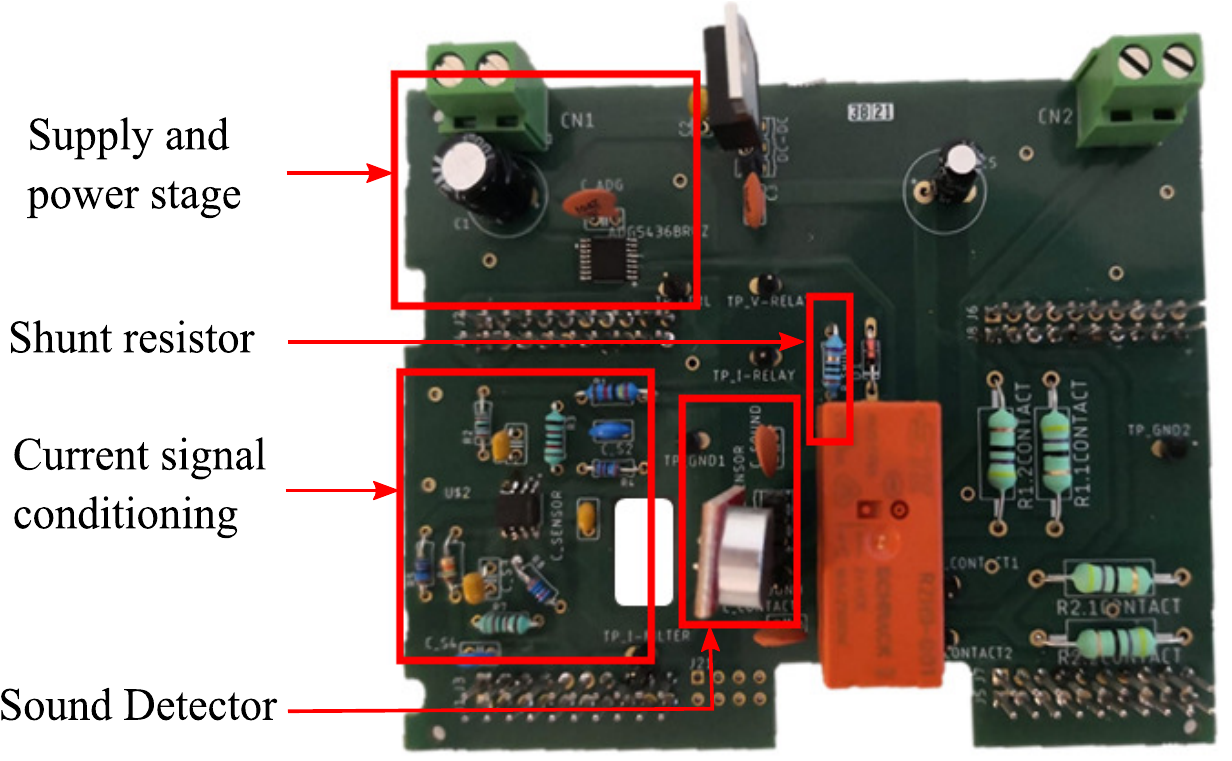}
	\caption{PCB Shield for adapting the signals between the microcontroller and the electromechancial relay (in orange).
	}
	\label{fig:shield}
\end{figure}

This subsection describes the experimental setup (Fig.~\ref{fig:environment}) used to analyze and validate the proposed control.
The entire control algorithm is implemented on a low-cost C2000 32-bit microcontroller and tested with various TE Connectivity RZ relays. It runs the feedforward and flux-tracking controllers in real time ($100\,\mathrm{kHz}$). On the other hand, the run-to-run adaptation law, which is executed between switching operations without real-time constraints, requires only a few milliseconds for each iteration.

A PCB shield (Fig.~\ref{fig:shield}) has been designed to adapt the signals from the microcontroller to the relay and vice versa. The shield includes an ADG5436 analog switch as power converter, which supplies a pulse-width modulated 
signal between $0\,\mathrm{V}$ and $35\,\mathrm{V}$ at $100\,\mathrm{kHz}$. It also includes a $10\,\mathrm{\Omega}$ shunt resistor for measuring the current, as well as its corresponding signal conditioning circuit. The current analog signal is filtered through a $16\,\mathrm{kHz}$ low-pass filter, whose amplitude is set to the microcontroller ADC reading range.
It consists of 2 operational amplifiers in Sallen-Key topology, which results in a $4$th order filter. A Bessel typology has been selected because this typology best preserves the signal shape.
The shield also includes a low-cost commercial sound detector (Sparkfun BOB-12758 Electret microphone) for measuring the impact noise.

The workbench is complemented by a Keyence laser sensor (LK-H082 sensor head and LK-G5001P controller). The measurement offered by this sensor is never used as control feedback, but only for preliminary estimation of nominal parameter values.
Lastly, all important signals are captured by an eight-channel USB oscilloscope (PicoScope 4824), and sent to a personal computer for storage and analysis.

\subsection{Experiments, results and discussion}\label{sec:results}

\begin{figure}
\centering
	\includegraphics{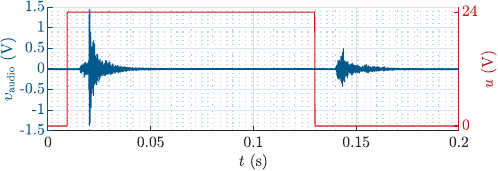}
	\caption{Sound signal in a conventional making and breaking switching cycle. The noise is higher during the making operation.}
	\label{fig:audio_switch}
\end{figure}

As previously stated, electromechanical relays perform two different operations: making and breaking. Although the presented controller is able to improve the performance of both, the switching sound is much higher in making operations (see Fig.~{\ref{fig:audio_switch}}). For this reason, the experimental analysis presented in this section is focused on the making operations. 

In order to initialize the control process, nominal parameter values need to be specified. The initial value of the parameter vector $p$ [see~{\eqref{eq:param0}}] has been obtained through an identification procedure, i.e., fitting the model to experimental measurements from one specific relay. Note, however, that the control process may also be initialized with arbitrarily chosen nominal parameters. Moreover, the desired trajectory time constants have been defined to closely match the switching times of a standard uncontrolled scenario ($0$ to $24\,\mathrm V$ square signal). The selected times are $\tc = 6.5\,\mathrm{ms}$ and $\tf = 8\,\mathrm{ms}$. Regarding the PI controller [see~{\eqref{eq:PI}}], the control gains have been preset so that the closed-loop response has a settling time (with a $5\,\%$ error band) less than or equal to $2$~ms and a damping coefficient greater than or equal to $1$. Then, they have been experimentally fine-tuned, resulting in the  parameter values $K_\mathrm{p} = 37500$~Wb/V and $K_\mathrm{i} = 1.15\times10^8$~Wb/(Vs).

In a real scenario, only one iterative control process would be needed to control any given relay.
Nonetheless, due to the high variability between and within relays, we have decided to conduct several trials in order to show the average performance and the variability of the control process. Specifically, the proposed control algorithm has been applied to 10 different relays and, for each one, 10 repetitions of the control process have been executed. Thus, 100 different trials have been performed, each one starting from the nominal parameter set. In addition to that, note also that in a real situation the controller would be working indefinitely, adapting its parameters each time the relay is switched. However, in the graphs in this section we only show the results obtained for the first 300 switching operations, which is a sufficient number to show the convergence of the method. In order to demonstrate the robustness to resistance variations, a $150\,\mathrm \Omega$ resistor has been added in series to the coil during the last 50 operations. This emulates a sudden temperature increase of $25\,^\circ$C, which would imply a 10\% increase in the copper electrical resistivity. Additionally, for comparison purposes, five standard switching operations with a constant $24$~V voltage signal have been carried out with no additional resistance, and another five standard operations with the $150\,\mathrm \Omega$ series resistor. Note that, in a practical scenario, environmental noise can negatively affect the control performance. Thus, for a fairer experimental validation, the tests have been carried out with no sound isolation. It is also worth remarking that the relays are enclosed and therefore a visual inspection of their moving parts is not possible. Thus, in order to maintain the integrity of the relays, the angular position has not been measured.

\begin{figure}[t]
    \begin{subfigure}{1\linewidth}
        \includegraphics{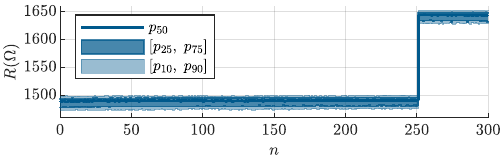}
        \caption{\label{fig:R_prctl}}
    \end{subfigure}%
    \vspace{\floatsep}
    \begin{subfigure}{1\linewidth}
	    \includegraphics{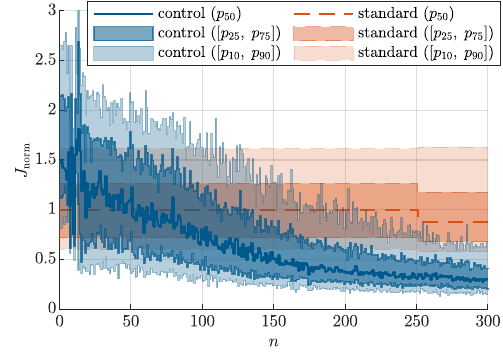}
	    \caption{\label{fig:y_prctl}}
	\end{subfigure}
	\caption{Distribution of values of the resistance (a) and normalized costs (b) for the 100 experimental trials as a function of the number of operations. Both graphs include the median value ($p_{50}$), the interquartile range ($[p_{25},\,p_{75}]$) and the $10$th to $90$th percentile interval ($[p_{10},\,p_{90}]$) of the corresponding distributions. The costs graph (b) shows the values obtained with control and with standard operations.}
\end{figure}

\begin{figure*}[!ht]
    \begin{subfigure}{0.332\linewidth}
        \centering
        \includegraphics[]{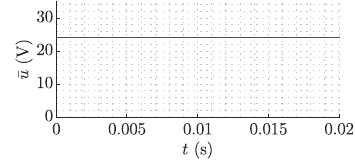}
        \caption{\label{fig:u_0}}
    \end{subfigure}%
    \begin{subfigure}{0.332\linewidth}
        \centering
        \includegraphics[]{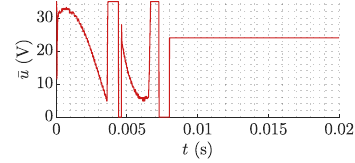}
        \caption{\label{fig:u_f1}}
    \end{subfigure}
    \begin{subfigure}{0.332\linewidth}
        \centering
        \includegraphics[]{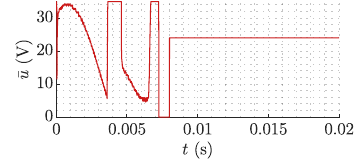}
        \caption{\label{fig:u_f2}}
    \end{subfigure}%
    \vspace{\floatsep}
    \begin{subfigure}{0.332\linewidth}
        \centering
        \includegraphics[]{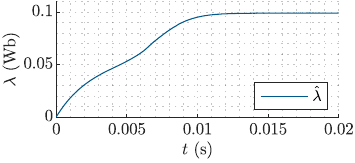}
        \caption{\label{fig:flux_0}}
    \end{subfigure}%
    \begin{subfigure}{0.332\linewidth}
        \centering
        \includegraphics[]{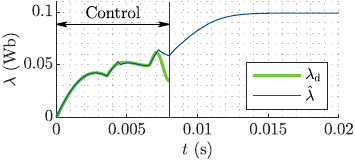}
        \caption{\label{fig:flux_f1}}
    \end{subfigure}
        \begin{subfigure}{0.332\linewidth}
        \centering
        \includegraphics[]{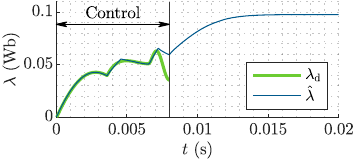}
        \caption{\label{fig:flux_f2}}
    \end{subfigure}%
    \vspace{\floatsep}
    \begin{subfigure}{0.332\linewidth}
        \centering
        \includegraphics[]{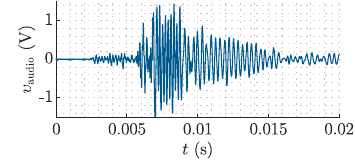}
        \caption{\label{fig:audio_0}}
    \end{subfigure}%
    \begin{subfigure}{0.332\linewidth}
        \centering
        \includegraphics[]{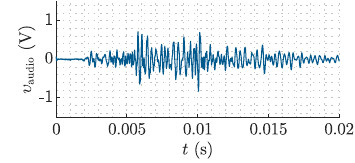}
        \caption{\label{fig:audio_f1}}
    \end{subfigure}
        \begin{subfigure}{0.332\linewidth}
        \centering
        \includegraphics[]{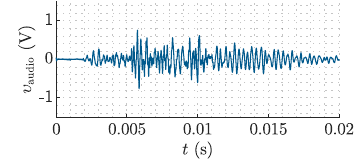}
        \caption{\label{fig:audio_f2}}
    \end{subfigure}
	\caption{Evolution of the control performance for a single relay. Voltage (a), flux linkage (d) and audio (g) in a standard operation.
	Voltage (b), flux linkage (e) and audio (h) in the last control iteration with low resistance ($n=250$). Voltage (c), flux linkage (f) and audio (i) in the first control iteration with increased resistance ($n=251$).}
	\label{fig:relay_results}
\end{figure*}

In the tests, the resistance value has been estimated for each of the 300 operations of the 100 performed experimental trials. The obtained distribution of values is represented in Fig.~{\ref{fig:R_prctl}}, by means of the $10$th, $25$th, $50$th, $75$th and $90$th percentiles ($p_{10}$, $p_{25}$, $p_{50}$, $p_{75}$ and $p_{90}$, respectively), as a function of the number of operations, $n \in \left[ 1\,.\,.\,300 \right]$. It is shown that there is some variability due to temperature fluctuations during the experimental testing, and due to differences between the coils of the tested relays. Nonetheless, the major resistance changes occur between $n=250$ and $n=251$, when the series resistor is connected.

Then, Fig.~{\ref{fig:y_prctl}} summarizes the control results. For a clear comparison, all the computed costs have been normalized with respect to the median cost of the standard switching operations at room temperature (i.e., without the series resistor). Equivalently to the resistance plot, this graph shows the distribution of values of the normalized cost, $J_\mathrm{norm}$, obtained for the 100 experimental trials performed, as a function of the number of control operations. For comparative purposes, it also displays the percentiles of the conventional switching operations without ($n \leq {250}$) and with ($n \geq {251}$) the series resistor.
The figure shows the high variability between the different tested devices, even though they are all of the same model. This variability is more evident at the beginning of the control process. Note that the costs are larger than in the uncontrolled scenarios, even though the control has been initialized using the estimated parameters from a similar relay. This shows the importance of the iterative adaptation of the model parameters, as explained in Section~{\ref{sec:r2r}}. Then, the control results improve greatly as the number of iterations increases. It is shown that the control requires about 50 iterations to improve the median with respect to standard operations. In a similar way, the control needs around 100 iterations for improving the $90$th percentile. The results also show that the control is robust against resistance variations, as the sudden resistance change does not seem to affect the normalized costs. In contrast, the standard operations are quite sensitive to temperature. In this regard, it can be seen that when the resistance increases, the median cost decreases and the $10$ to $90$th percentile interval widens.

As a more intuitive demonstration of the control performance and its robustness against resistance variations, a comparison of voltage, flux linkage and audio signals is presented in Fig.~{\ref{fig:relay_results}}. 
These results correspond to the trial with the largest audio costs in the standard uncontrolled scenario. Figs.~{\ref{fig:u_0}}, {\ref{fig:flux_0}} and~{\ref{fig:audio_0}} represent the voltage applied in a standard operation, the estimated flux linkage and the resulting audio signal, respectively. Then, Figs.~{\ref{fig:u_f1}}, {\ref{fig:flux_f1}} and~{\ref{fig:audio_f1}} show the same signals for the 250th control iteration. As can be seen in Fig.~{\ref{fig:flux_f1}}, the flux-tracking control works correctly. The tracking only fails at the end of the control interval, i.e., close to $8$~ms, but this is due to voltage saturation (see Fig.~{\ref{fig:u_f1}}). Despite the voltage signal limitations of the implementation, it manages to reduce considerably the audio intensity, which indicates that the switching impacts are also decreased. Then, Figs.~{\ref{fig:u_f2}}, {\ref{fig:flux_f2}} and~{\ref{fig:audio_f2}} correspond to the 251th control iteration, in which the resistance increases by $150\,\mathrm \Omega$. Due to this change, the resulting voltage signal is modified (see Fig.~{\ref{fig:u_f2}}). However, as expected, the desired flux is unaffected by this change (see Fig.~{\ref{fig:flux_f2}}). Furthermore, the flux-tracking control performs very similarly despite the sudden change in its dynamics and, consequently, the resulting noise reduction is almost identical (see Fig.~{\ref{fig:audio_f2}}).

\subsection{Further results: flux-based vs. voltage-based control}\label{subsec:volt_control}

As explained in Section~{\ref{sec:control}}, the proposed flux-based feedforward controller is advantageous over the current- and voltage-based counterparts because it depends on fewer parameters. To complement the theoretical explanation, this subsection includes an experimental comparison of our proposal with an alternative voltage-based controller. This controller is based on the same principles as the other, but the feedforward block calculates directly the input voltage from the desired trajectory as described in~{\eqref{eq:u_flatness}}. The flux-tracking controller and estimator from Section~{\ref{sec:flux_control}} are thus not used in this case. Nonetheless, note that the feedforward controller still uses the resistance estimation in order to calculate the voltage signal.

\begin{figure}[t]
	\includegraphics{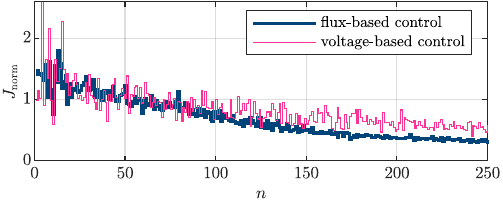}
	\caption{Median values of the normalized costs for the 100 experimental trials as a function of the number of operations. Comparison between the flux-based controller and the alternative voltage-based controller.}
	\label{fig:y_volt}
\end{figure}

The alternative controller has been experimentally tested in the same manner and using the same ten relays as in the previous subsection for the first 250 iterations (i.e., with no series resistor). The obtained results are compared in Fig.~{\ref{fig:y_volt}} with the ones from the flux-based controller. For clarity reasons, the graphic depicts only the median costs $J_\mathrm{norm}$ of the controllers instead of the complete distributions of values. There is an initial phase in which the run-to-run adaptation law performs a broad search in the parameter space. This can be seen in the fact that the costs obtained with any of the two controllers vary greatly in the first 20 operations. For the following operations, both controllers start to converge, resulting in decreasing costs. Note that, although the results are initially similar, there is a larger variability between consecutive operations for the voltage-based controller. Then, for the last operations ($n\geq 150$), the most noticeable aspect is that the average costs from the flux-based feedforward controller are consistently smaller, which confirms that the proposed design performs better than the voltage-based version.

\section{Conclusion}\label{sec:con}

In this paper we have proposed an audio-based control strategy to reduce the negative effects of impacts during relay switching operations, accounting for the nonlinear dynamics and high variability of these devices. The experimental results show that this control proposal is able to reduce not only the average impact noise, but also its variability. 

The proposal includes flux-based feedforward and feedback control blocks, which have several advantages over voltage- or current-based ones. Most importantly, they make the control more robust against modeling errors and temperature variations. Additionally, an adaptive learning-type run-to-run block is incorporated to circumvent modeling and other errors and to improve the control performance. This is specially useful for devices with high variability between operations or between units, as the studied case. This proposed run-to-run adaptation law relies on microphone measurements, but, in a more general case, they can be easily replaced or complemented with auxiliary measurements, e.g. the contact voltages for detecting bounces. Moreover, the modular control structure is very versatile, permitting the replacement or modification of any of the control blocks.

\appendix

\subsection{Stability proof of the flux-tracking controller}\label{appendix:stability}

As stated, the flux linkage dynamics {\eqref{eq:mag_circuit}} is controlled using a PI controller in parallel form. Let the controller equation {\eqref{eq:PI}} be reformulated as
\begin{align}
    u(t) &= K_\mathrm{p} \left(\lambda_\mathrm{d}(t)-\hat\lambda(t)\right) + K_\mathrm{i}\,\sigma(t), \label{eq:PI2}\\
    \dot \sigma(t) &= \lambda_\mathrm{d}(t)-\hat\lambda(t), \label{eq:PI3}
\end{align}
where $\sigma$ is the integral of the error. Then, using {\eqref{eq:mag_circuit}}, {\eqref{eq:PI2}} and {\eqref{eq:PI3}}, it is possible to describe the closed-loop dynamics as
\begin{multline}
    \left[\begin{array}{c}
        \dot \lambda(t) \\
        \dot \sigma(t)
        \end{array}\right] = 
        \left[\begin{array}{cc}
        - a(t)-K_\mathrm{p} & K_\mathrm{i} \\
        -1 & 0
    \end{array}\right]
    \left[\begin{array}{c}
        \lambda(t) \\
        \sigma(t)
    \end{array}\right] \\
    +
    \left[\begin{array}{cc}
        K_\mathrm{p} & K_\mathrm{p} \\
        1 & 1
        \end{array}\right]
    \left[\begin{array}{c}
        \lambdaref(t) \\
        \tilde\lambda(t)
    \end{array}\right],
\end{multline}
where $\tilde\lambda=\lambda-\hat\lambda$ is the estimation error. The eigenvalues of the closed-loop system matrix are given by
\begin{multline}
    \mathrm{eig}    
    \left(\left[\begin{array}{cc}
    - a(t)-K_\mathrm{p} & K_\mathrm{i} \\
    -1 & 0
    \end{array}\right]\right) \\
    = \dfrac{-a(t) - K_\mathrm{p} \pm \sqrt{\left(a(t)+K_\mathrm{p}\right)^2-4\,K_\mathrm{i}}}{2}.
\end{multline}
Since $a(t)>0$ for all $t$, the eigenvalues of the system will have negative real part for any positive value of $K_\mathrm{p}$ and $K_\mathrm{i}$. Thus, it can be concluded that the closed-loop system is stable if $K_\mathrm{p}>0$ and $K_\mathrm{i}>0$. The steady state regime is given by
\begin{align}
    \left[\begin{array}{c}
        \lambda(t) \\
        \sigma(t)
    \end{array}\right] =
    \left[\begin{array}{cc}
        1 & 1 \\
        a(t)/K_\mathrm{i} & a(t)/K_\mathrm{i}
    \end{array}\right]
    \left[\begin{array}{c}
        \lambdaref(t) \\
        \tilde\lambda(t)
    \end{array}\right],
\end{align}
which shows that $\lambda$ converges to the desired value $\lambdaref$ provided that the estimation error $\tilde\lambda$ converges to zero.

\end{document}